\documentclass[12pt,preprint]{aastex}

\shorttitle{An off-centered AGN in NGC 3115}
\shortauthors{Menezes et al.}

\begin{document}

\title{An off-centered active galactic nucleus in NGC 3115}

\author{R. B. Menezes, J. E. Steiner, and T. V. Ricci}

\affil{Instituto de Astronomia Geof\'isica e Ci\^encias Atmosf\'ericas, Universidade de S\~ao Paulo, Rua do Mat\~ao 1226, Cidade Universit\'aria, S\~ao Paulo, SP CEP 05508-090, Brazil;}
\email{robertobm@astro.iag.usp.br}

\begin{abstract}

NGC 3115 is an S0 galaxy that has always been considered to have a pure absorption-line spectrum. Some recent studies have detected a compact radio-emitting nucleus in this object, coinciding with the photometric center and with a candidate for the X-ray nucleus. This is evidence of the existence of a low-luminosity active galactic nucleus (AGN) in the galaxy, although no emission line has ever been observed. We report the detection of an emission-line spectrum of a type 1 AGN in NGC 3115, with an H$\alpha$ luminosity of $L_{H\alpha} = (4.2 \pm 0.4) \times 10^{37}$ erg s$^{-1}$. Our analysis revealed that this AGN is located at a projected distance of $\sim 0\arcsec\!\!.29 \pm 0\arcsec\!\!.05$ (corresponding to $\sim 14.3 \pm 2.5$ pc) from the stellar bulge center, which is coincident with the kinematic center of this object's stellar velocity map. The black hole corresponding to the observed off-centered AGN may form a binary system with a black hole located at the stellar bulge center. However, it is also possible that the displaced black hole is the merged remnant of the binary system coalescence, after the ``kick'' caused by the asymmetric emission of gravitational waves. We propose that certain features in the stellar velocity dispersion map are the result of perturbations caused by the off-centered AGN.

\end{abstract}


\section{Introduction}

NGC 3115 is an almost edge-on  S0 galaxy at a distance of about 10.2 Mpc. This object has always been considered to have a typical absorption spectrum without emission lines. For this reason, \citet{ho97} used this galaxy's spectrum as a template for starlight subtraction from the spectra of other galaxies. However, \citet{wro12}, using 8.5 GHz observations obtained with the Very Large Array, detected a radio-emitting nucleus in NGC 3115, with a diameter smaller than $0\arcsec\!\!.17$ (8.4 pc) and luminosity in 8.5 GHz of $3.1 \times 10^{35}$ erg s$^{-1}$. This radio source coincides with the photometric center of the galaxy and also with a candidate for the X-ray nucleus. Therefore, the detection made by \citet{wro12} is evidence of a low-luminosity active galactic nucleus (AGN) in this object, although no emission line has been reported so far. \citet{won11}, using X-ray observations made with the \textit{Chandra X-ray Observatory}, resolved, for the first time, the accretion flux of gas toward the central black hole in NGC 3115 and also estimated that the accretion rate, at the Bondi radius, is approximately $2.2 \times 10^{-2} M_{\sun}$ yr$^{-1}$. 

In this work, we report the detection, for the first time, of emission lines indicating the existence of an off-centered type 1 AGN in NGC 3115.

\section{Observations, reduction and data treatment}

The observations of NGC 3115 were made with the integral field unity (IFU) of the Gemini Multi-Object Spectrograph (GMOS), mounted on the Gemini-South telescope, on 2013 June 30. We obtained one 30 minute exposure, with the science field of view centered on the nuclear region of NGC 3115. We used the B600-G5323 grating, in a central wavelength of $5620\AA$, which provides a spectral coverage of $4185\AA - 7040\AA$ and a resolution of $R \sim 3900$. Based on the acquisition images, we estimated a seeing of $\sim 0\arcsec\!\!.8$ for the night of observation. 

The data reduction was performed with the Gemini IRAF package. This process resulted in one data cube, with spatial pixels (spaxels) of $0\arcsec\!\!.05 \times 0\arcsec\!\!.05$. These spaxels were artificially created, without any smoothing, during the data cube construction. The original size of the spaxels of the GMOS-IFU is $0\arcsec\!\!.2 \times 0\arcsec\!\!.2$. We verified that this spatial re-sampling usually results in an uncertainty of $\sim 0\arcsec\!\!.02$ for the positions of the objects. This value was taken into account in the analysis.

\begin{figure}
\plotone{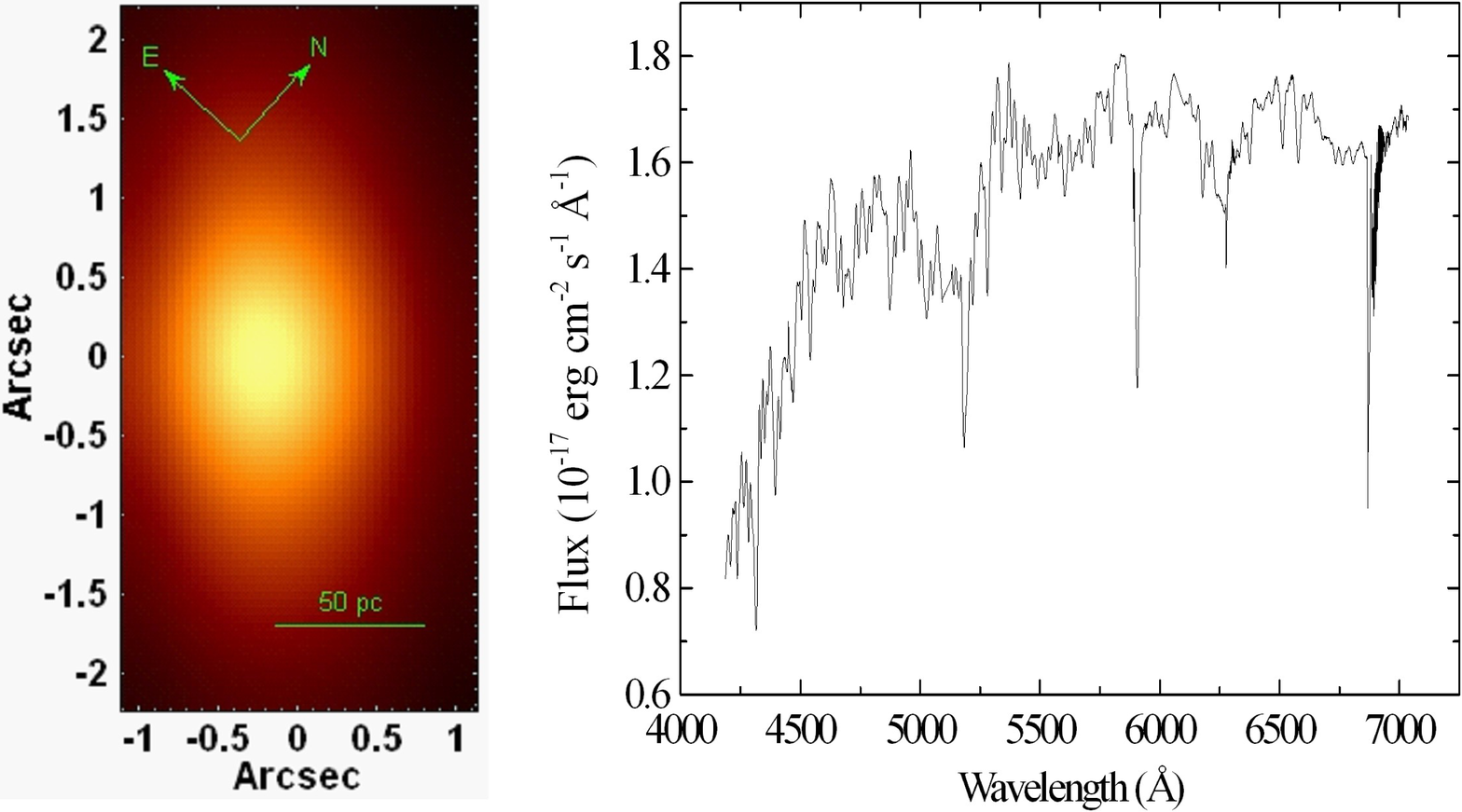}
\caption{Left: image of the final data cube of NGC 3115 collapsed along the spectral axis. Right: average spectrum of the final data cube of NGC 3115.\label{fig1}}
\end{figure}

We applied a treatment procedure to the reduced data cube, including the following steps: correction of the differential atmospheric refraction, Butterworth spatial filtering \citep{gon02}, instrumental fingerprint removal, and Richardson-Lucy deconvolution \citep{ric72, luc74}. The purpose of the Butterworth spatial filtering is to remove high spatial-frequency components from the images. Any uncertainty in the positions of the objects introduced by this filtering is significantly small ($\sim 0\arcsec\!\!.001$) and therefore was not taken into account in the analysis. In the instrumental fingerprint removal, we isolate and remove certain structures from the data cube, which appear as vertical stripes in the images and have a characteristic spectral signature. This procedure is performed with the PCA Tomography technique \citep{ste09}. Finally, we apply the Richardson-Lucy deconvolution, using a Gaussian point-spread function (PSF), in order to improve the spatial resolution of the data cube. The FWHM of the PSF of the final data cube is $\sim 0\arcsec\!\!.6$. For more details about the methodologies we used in the data treatment, see \citet{men14}. The average spectrum and the image of the treated data cube of NGC 3115, collapsed along the spectral axis, are shown in Figure~\ref{fig1}.

\section{Data analysis and results}

After the data treatment, we applied a correction of the interstellar extinction, due to the Galaxy, to the data cube, using $A_V = 0.127$ (NASA Extragalactic Database - NED) and the reddening law of \citet{car89}. We then shifted the spectra to the rest-frame, using $z = 0.002212$ (NED), and sampled them with $\Delta\lambda = 1\AA$. After this preparation, we performed a spectral synthesis to fit all the stellar spectra of the data cube. This procedure fits the observed spectra using a linear combination of stellar population spectra from a given base. One of the results was a synthetic stellar spectrum for each spaxel of the data cube. We applied the spectral synthesis with the {\sc starlight} software \citep{cid05}, using a base of stellar populations based on Medium-resolution Isaac Newton Telescope Library of Empirical Spectra (MILES; S\'anchez-Bl\'azquez et al. 2006), due to the similarity between its spectral resolution (FWHM = $2.3\AA$) and our spectral resolution (FWHM = $1.6 \AA$). 

\begin{figure}
\epsscale{0.63}
\plotone{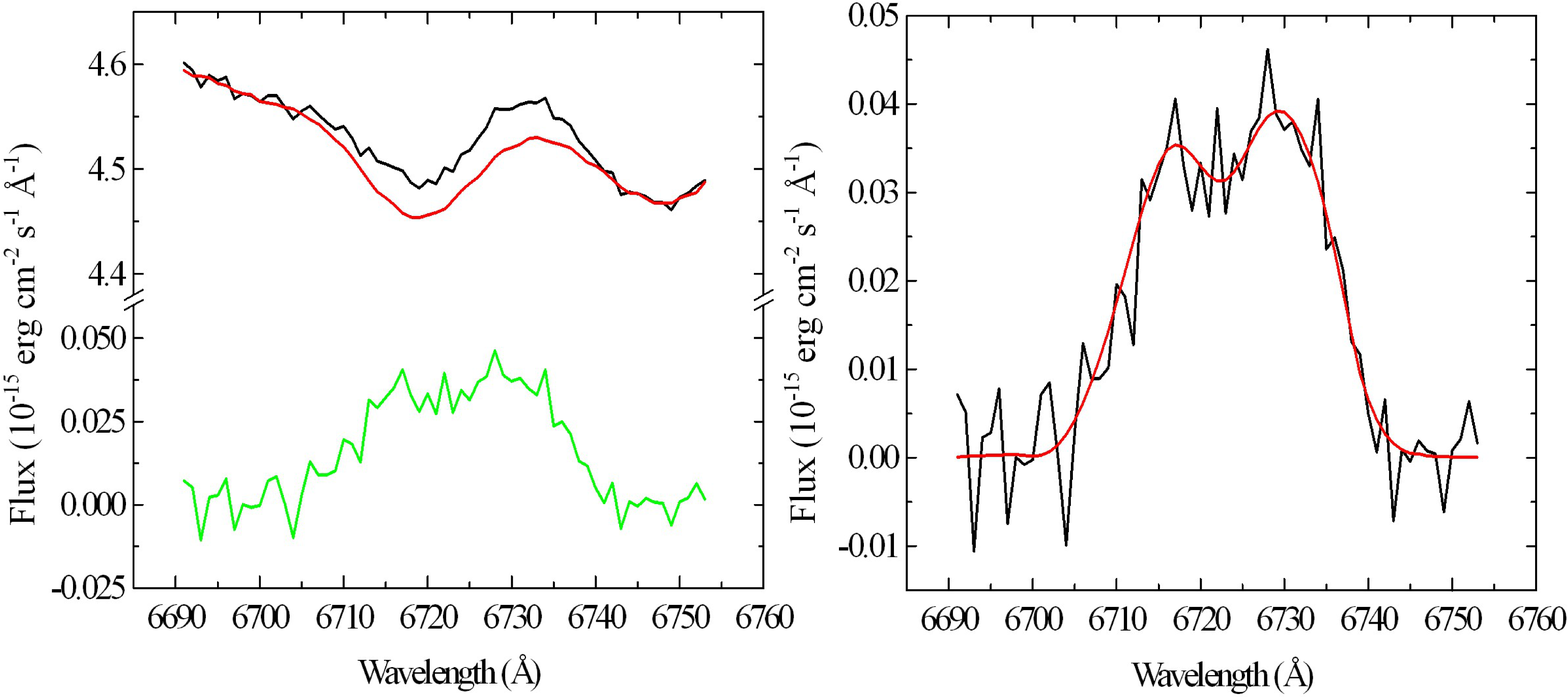}
\epsscale{0.61}
\plotone{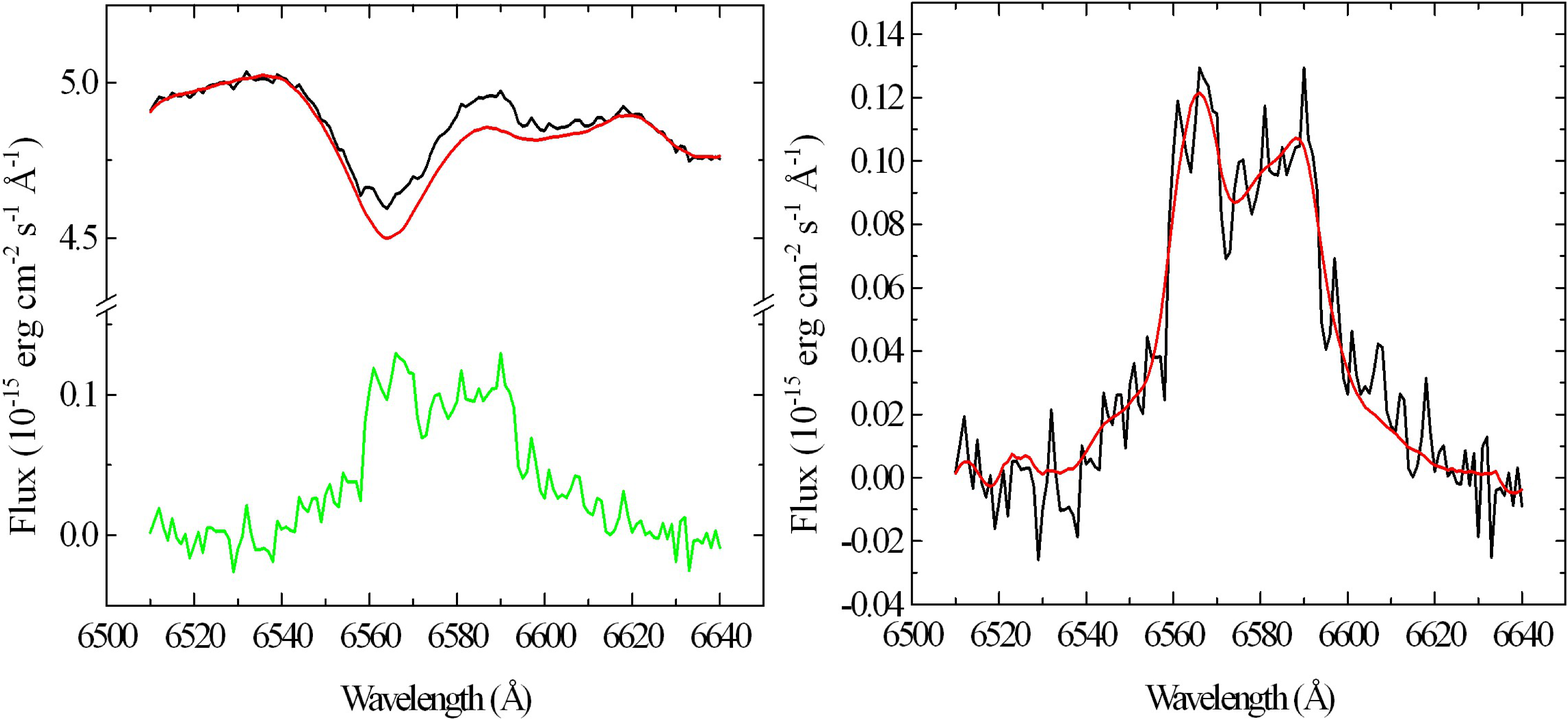}
\plotone{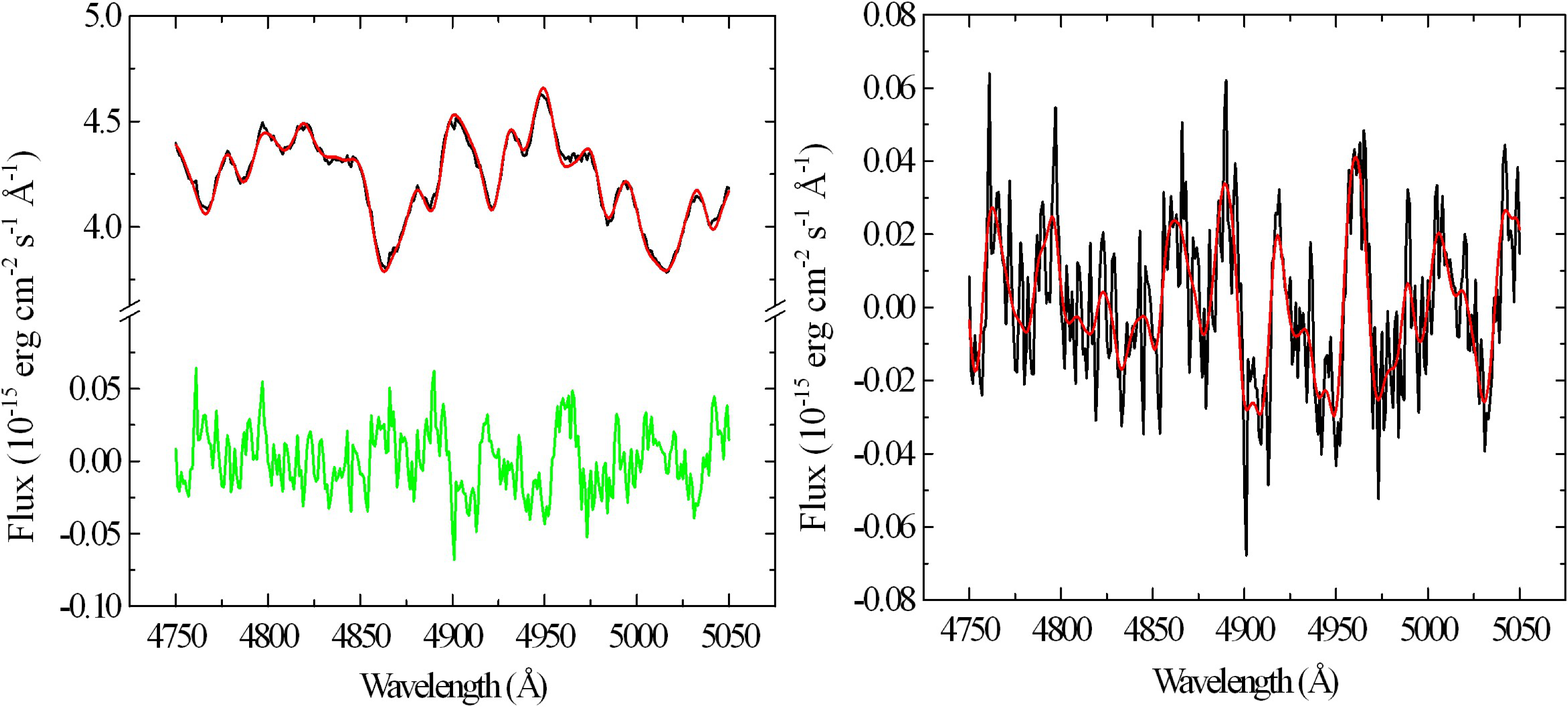}
\epsscale{0.18}\\
\plotone{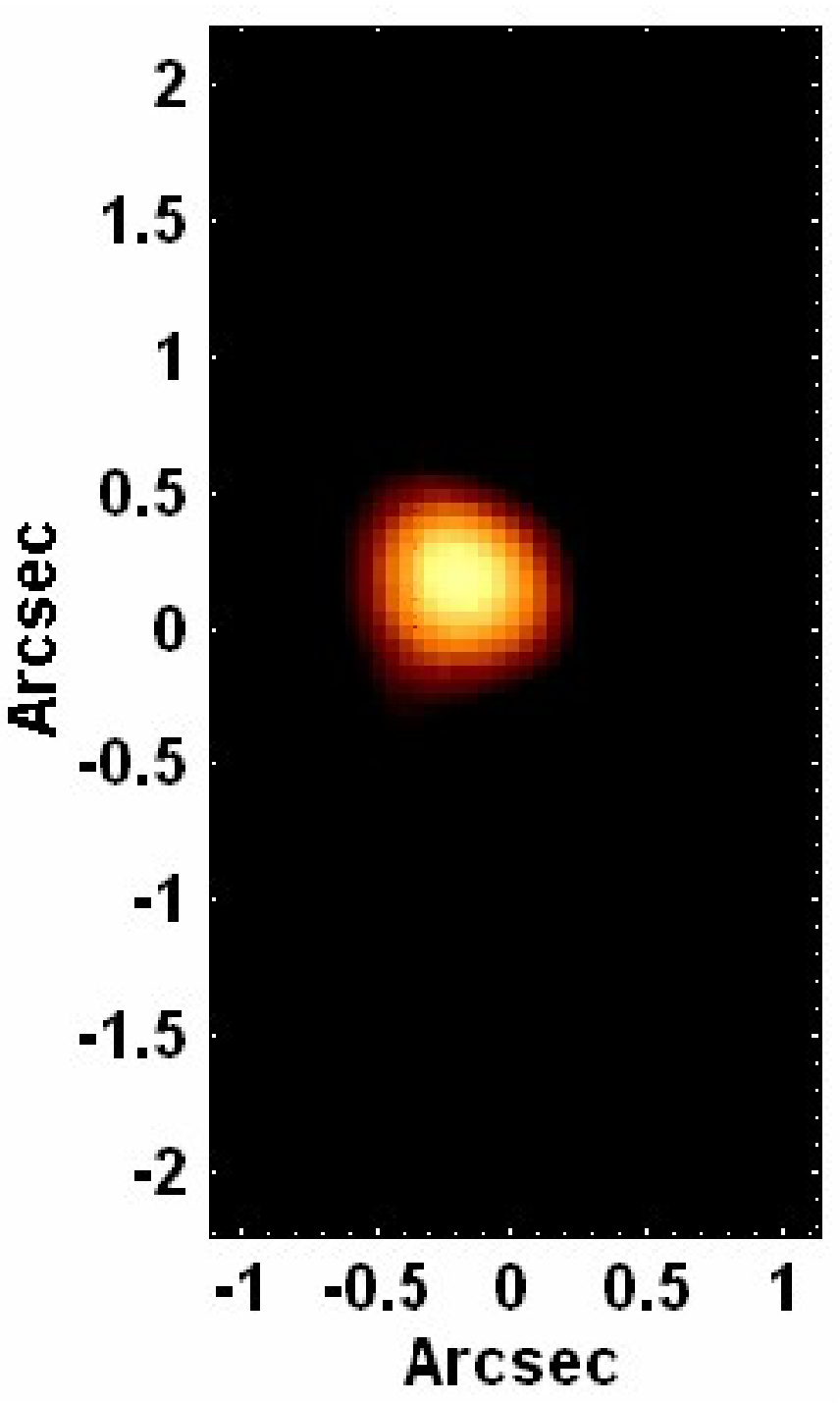}
\caption{Left: different wavelength ranges of the spectrum of a circular region, centered on the compact emitting region of the data cube of NGC 3115, with a radius of $0\arcsec\!\!.35$, extracted before starlight subtraction. The fit provided by the spectral synthesis is shown in red and the fit residuals are shown in green. Right: the same fit residuals shown at left, with the result obtained after the Butterworth spectral filtering shown in red. Bottom: image of the collapsed wavelength interval $6537 - 6607 \AA$ (which contains the H$\alpha$ and [N II] $\lambda \lambda6548,6583$ emission lines).\label{fig2}}
\end{figure}

The subtraction of the synthetic spectra from the observed ones resulted in a data cube containing only emission lines. In this subtracted data cube, we detected, for the first time, very weak emission lines. We constructed an image of the collapsed wavelength interval $6537 - 6607 \AA$ (which contains the H$\alpha$ and [N II] $\lambda \lambda 6548, 6583$ emission lines). The result, shown in Figure~\ref{fig2}, reveals that the observed emission lines are emitted in a very compact region. Figure~\ref{fig2} shows different wavelength ranges of the spectrum of a circular region, centered on the compact emitting region, with a radius of $0\arcsec\!\!.35$, extracted from the data cube of NGC 3115 before starlight subtraction. For each one of these wavelength ranges, we also show the fit provided by the spectral synthesis and the fit residuals. We can see the existence of peaks whose wavelengths are compatible with the wavelengths of the  H$\alpha$, [N II] $\lambda 6583$, and [S II] $\lambda \lambda 6716, 6731$ emission lines. There is also a more diffuse structure whose wavelength is compatible with the wavelength of the [N II] $\lambda 6548$ emission line. We observed two additional peaks with wavelengths of $\sim 4862 \AA$ and $\sim 5007 \AA$ that could be identified as the H$\beta$ and [O III] $\lambda 5007$ emission lines. However, since the intensities of these two peaks are comparable with the intensities of the fluctuations in adjacent spectral regions, we cannot confirm the detection of these two emission lines.

A possible broad component of the H$\alpha$ emission line can also be seen in the fit residuals in Figure~\ref{fig2}. The H$\alpha$ and [N II] emission lines, together with a possible broad component of H$\alpha$, suggest the presence of a type 1 AGN in NGC 3115. In order to fit the emission lines with sums of Gaussian functions, first of all, we removed most of the spectral noise with a Butterworth spectral filtering. The results of this procedure, shown in Figure~\ref{fig2}, reveal that the filtering successfully removed, in all spectral ranges analyzed here, the high-frequency noise without introducing artifacts or altering the width and the profile of the emission lines (this procedure also did not allow a clear detection of the  H$\beta$ and [O III] $\lambda 5007$ emission lines). The values of the signal-to-noise ratio in the spectral range $6561 \AA - 6591 \AA$ before and after the spectral filtering were $\sim 6.9$ and $\sim 17.0$, respectively.

\begin{figure}
\epsscale{0.8}
\plotone{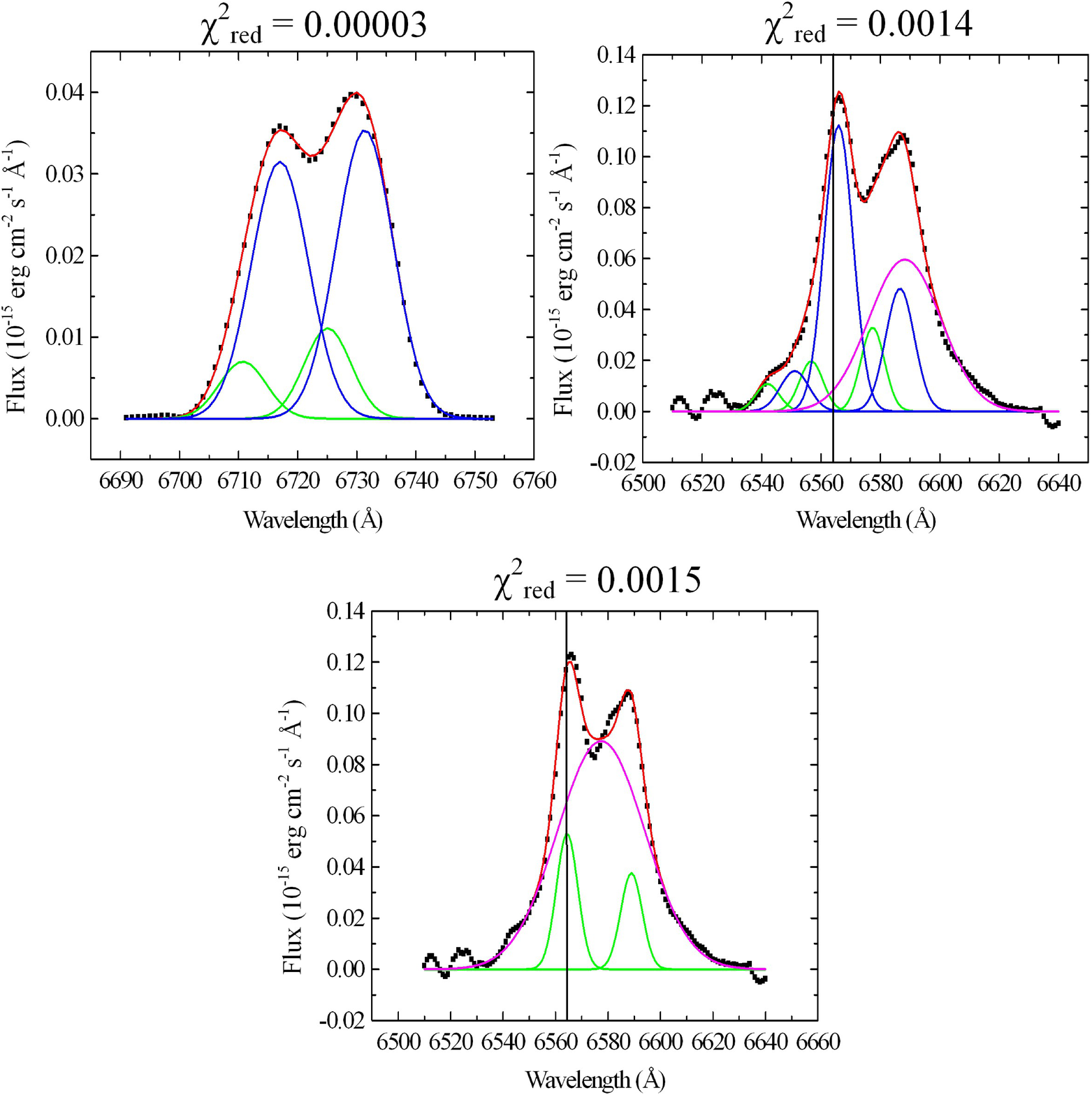}
\caption{Upper left: Gaussian fits applied to the [S II] region of the filtered residuals of the spectrum shown in Figure~\ref{fig2}. Upper right: Gaussian fits applied to the H$\alpha$ region of the same filtered residuals, taking into account the narrow components of the H$\alpha$ and [N II] $\lambda \lambda6548,6583$ emission lines and also the broad component of the H$\alpha$ line. Bottom: Gaussian fits applied to the same filtered residuals, assuming the existence of a double peak and of a broad component of the H$\alpha$ emission line. In all panels, the residuals correspond to the black points, the Gaussians in blue and green represent the narrow components of the emission lines, the Gaussian in magenta represents the broad component of H$\alpha$, and the final fits are shown in red. The vertical black line represents the wavelength of the H$\alpha$ absorption line in the spectrum before the starlight subtraction.\label{fig3}}
\end{figure}

After the spectral filtering, we fitted the [S II] $\lambda \lambda 6716, 6731$ emission lines with a sum of Gaussian functions. We used two sets of two narrow Gaussians, assuming one radial velocity and one width for each one of the sets. After this procedure, we fitted the H$\alpha$ and [N II] $\lambda \lambda6548,6583$ emission lines using two sets of three narrow Gaussians. However, in this case, we assumed that each one of these sets have the same radial velocity and width of the corresponding set used to fit the [S II] $\lambda \lambda 6716, 6731$ emission lines. In other words, we used the [S II] $\lambda \lambda 6716, 6731$ emission lines as an empirical template to fit the narrow components of the H$\alpha$ and [N II] $\lambda \lambda6548,6583$ emission lines \citep{ho97b}. To fit the possible broad component of the H$\alpha$ line, we added a broader Gaussian function to the fit. The relative intensities between the Gaussians used to fit the [N II] $\lambda6548$ and [N II] $\lambda6583$ lines were kept constant, as established theoretically \citep{ost06}. The results, together with the reduced $\chi^2$s (which correspond to the $\chi^2$s divided by the number of degrees of freedom), are shown in Figure~\ref{fig3}. The fits reproduced, with good precision, the observed emission lines and the existence of a broad component of H$\alpha$ was confirmed. The values of the FWHM of the narrow Gaussians used to fit the emission lines are $\sim 421$ km s$^{-1}$ and $\sim 515$ km s$^{-1}$. On the other hand, the broad component of H$\alpha$ has a FWHM of $\sim 1307$ km s$^{-1}$. This broad component is displaced to the red, with a radial velocity of $V_{H\alpha}(broad) \sim1162$ km s$^{-1}$.

Using the Gaussian fits shown in Figure~\ref{fig3}, we calculated the [N II] $\lambda6584$/H$\alpha$ emission-line ratio and obtained a value of $0.59 \pm 0.08$, which is characteristic of AGNs \citep{bal81}. We also obtained an H$\alpha$ luminosity of $L_{H\alpha} = (4.2 \pm 0.4) \times 10^{37}$ erg s$^{-1}$. Finally, using the relation $L_{bol} \approx 220L_{H\alpha}$ \citep{ho08}, we estimated the bolometric luminosity of the AGN and obtained a value of $L_{bol} = (9.2 \pm 0.9) \times 10^{39}$ erg s$^{-1}$.

A different scenario may also be proposed: the two peaks with the highest fluxes in the emission-line spectrum in Figures 2 and 3 could be interpreted as the double peak of the H$\alpha$ emission line, generated by a rotating relativistic disk around the central black hole in NGC 3115. In order to evaluate this hypothesis, we fitted the observed emission lines with a sum of two narrow Gaussians (with the same width) and one broad Gaussian. The quality of the obtained fit, shown in Figure~\ref{fig3}, is similar to the quality of the fit obtained assuming two sets of three narrow Gaussians and one broad Gaussian. The FWHM of the narrow Gaussians in this new fit is $\sim 433$ km s$^{-1}$, while the FWHM of the broad component of H$\alpha$ is $\sim 1789$ km s$^{-1}$. The hypothesis of the existence of a rotating relativistic disk will be discussed in further detail below.

\begin{figure}
\epsscale{1.0}
\plotone{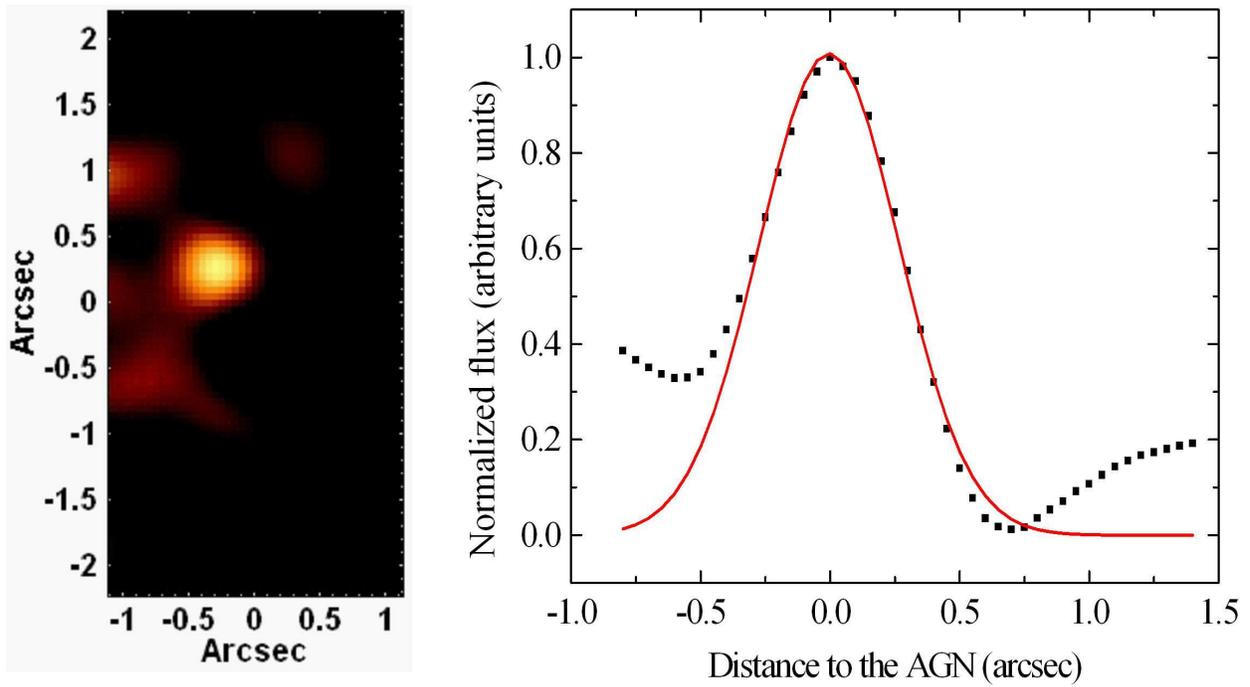}
\caption{Left: image of the collapsed wavelength interval $6595 - 6613 \AA$, which corresponds to the red wing of the broad component of H$\alpha$. Right: normalized horizontal brightness profile of the image shown at left, with a Gaussian fit shown in red.\label{fig4}}
\end{figure}

A natural question at this point is: could the emission lines in the spectrum of Figures 2 and 3 be simply the result of imprecise starlight subtraction ? If this was the case, we would expect an image of the collapsed wavelength interval containing the observed emission lines to reveal basically noise, not well-defined emitting areas, like the image of the collapsed wavelength interval $6537 - 6607 \AA$ in Figure~\ref{fig2} shows. Besides that, we verified that the use of different bases of stellar populations \citep{bru03,wal09} in the spectral synthesis does not qualitatively affect the emission-line spectra detected after the starlight subtraction (but implies an uncertainty of about 10\% in the flux of the emission lines). In order to make a more detailed analysis of this topic, we constructed an image of the red wing of the broad component of H$\alpha$ (in the wavelength range $6595 - 6613 \AA$). We also extracted the normalized horizontal brightness profile of the obtained image. The results are shown in Figure~\ref{fig4} and reveal that the broad component of H$\alpha$ is emitted essentially in a point-like region. This is confirmed by the fact that the Gaussian fit applied to the normalized horizontal brightness profile has a FWHM of $\sim 0\arcsec\!\!.63$, which is very similar to the estimated FWHM of the PSF of the treated data cube of NGC 3115 ($\sim 0\arcsec\!\!.6$). At the distance of typical AGNs, the broad-line region, which is the source of the broad components in permitted lines, can be regarded as a point-like source. Therefore, the fact that the image of the red wing of the broad component of H$\alpha$ reveals a point-like emitting area confirms not only that the observed emission lines are not the result of imprecise starlight subtraction, but also that the emission comes from a type 1 AGN.

We applied the penalized pixel fitting (pPXF) method \citep{cap04} to the spectrum of each spaxel of the data cube of NGC 3115 before starlight subtraction. This procedure fits the stellar spectrum of a given object using a Gauss-Hermite expansion and a combination of template stellar population spectra from a pre-established base. We once again used the base of stellar populations based on MILES. The pPXF method provides the values of the following parameters: stellar radial velocity ($V_*$), stellar velocity dispersion ($\sigma_*$), and the Gauss-Hermite coefficients $h_3$ and $h_4$. Figure~\ref{fig5} shows the maps of $V_*$ and $\sigma_*$. The uncertainties of the values of $V_*$ and $\sigma_*$ (obtained using Monte Carlo simulations) are in the ranges of $3-8$ km s$^{-1}$ and $5-10$ km s$^{-1}$, respectively. We constructed an RGB composite image with the image of the data cube (before starlight subtraction) collapsed along the spectral axis shown in green, the image of the red wing of the broad component of H$\alpha$ shown in red, and the $\sigma_*$ map shown in blue. The result can be seen in Figure~\ref{fig5}. The kinematic center of NGC 3115, which was defined as the point along the line of nodes of the $V_*$ map with velocity equal to zero, is shown in this RGB with a black cross. Considering that the image of the collapsed data cube essentially represents the central region of the stellar bulge of NGC 3115, we can say that the RGB composite image in Figure~\ref{fig5} makes it clear that the position of the AGN does not coincide with the stellar bulge center. The projected distance between the AGN and the center of the bulge is $\sim 0\arcsec\!\!.29 \pm 0\arcsec\!\!.05$ (corresponding to $\sim 14.3 \pm 2.5$ pc). The uncertainty of this value was obtained using a Monte Carlo simulation and also taking into account the uncertainty introduced by the spatial re-sampling during the data reduction. The RGB in Figure~\ref{fig5} also reveals that the kinematic center is compatible with the stellar bulge center, but not with the $\sigma_*$ peak, which is also not coincident with the position of the AGN.

\section{Discussion and conclusions}

The detection of an emission-line spectrum of a type 1 AGN in NGC 3115 is compatible with the detection of a compact radio-emitting nucleus in this galaxy \citep{wro12}. \citet{zha09} identified a point-like \textit{Chandra} source close to the photometric center of NGC 3115; however, \citet{won14}, also using \textit{Chandra} data, found no strong evidence of a nuclear point-like source, concluding that the central peak is extended, and estimated a very low upper limit for the X-ray luminosity of an AGN in this galaxy ($L_X < 4.4 (1.1) \times 10^{37}$ erg s$^{-1}$ in 0.5-6.0 (0.5-1.0) keV). Similarly, the value of $L_{bol} = (9.2 \pm 0.9) \times 10^{39}$ erg s$^{-1}$ that we determined for the AGN is quite low, both for low-ionization nuclear emission-line regions and Seyferts \citep{ho08}.

The hypothesis that the observed emission-line spectrum of NGC 3115 comprises the H$\alpha$ and [N II] $\lambda \lambda6548,6583$ narrow lines and also a broad component of H$\alpha$ is certainly more likely than the scenario involving the existence of a rotating relativistic disk around the central black hole. One reason is the existence of three structures whose wavelengths are compatible with the wavelengths of the H$\alpha$ and [N II] $\lambda \lambda6548,6583$ emission lines. Besides that, in the relativistic disk scenario, the two peaks with highest fluxes would represent the double peak of the H$\alpha$ emission line and, as a consequence, there would not be [N II] emission lines, which is unlikely in this case.

\begin{figure}
\epsscale{0.9}
\plotone{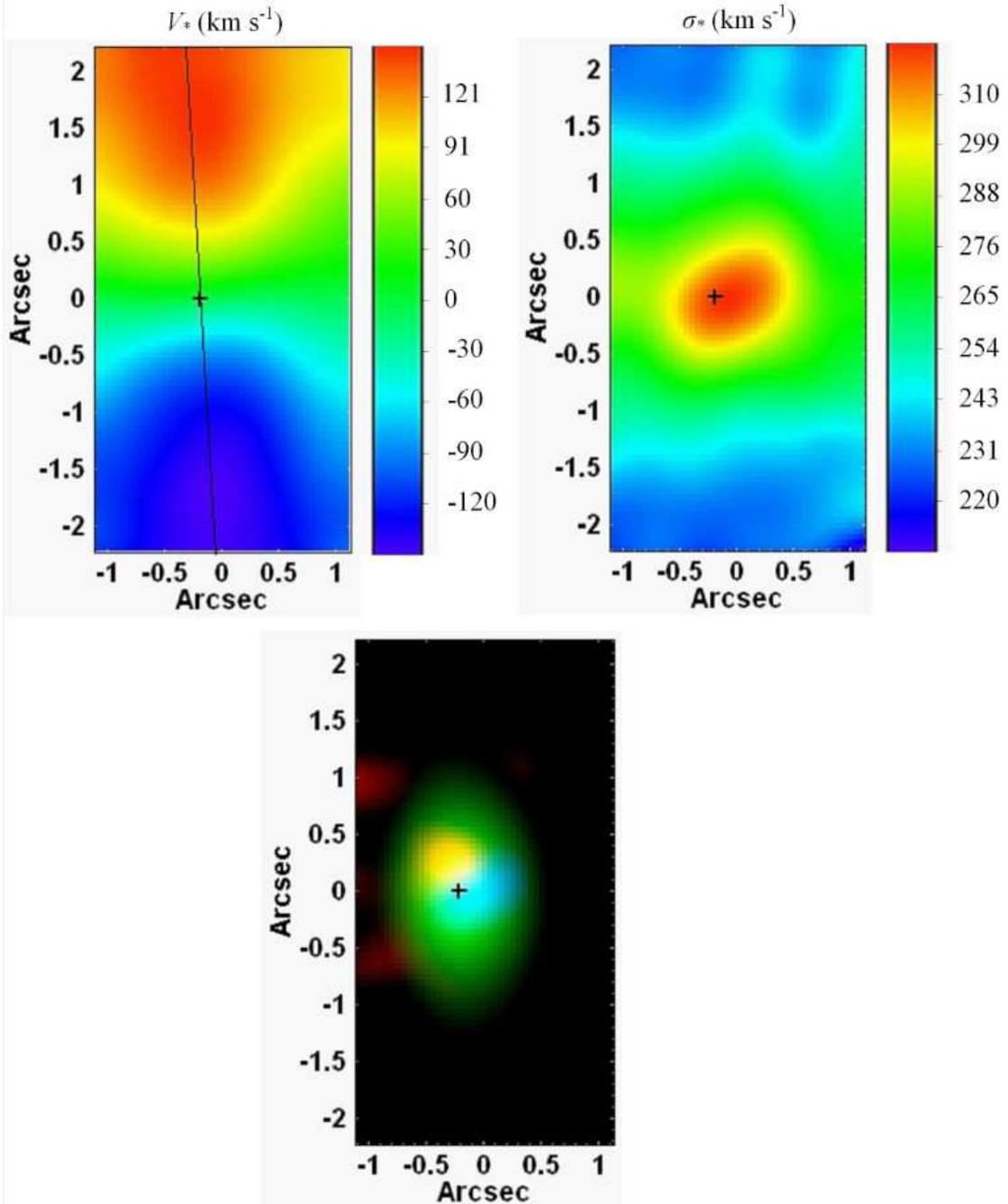}
\caption{Upper left: $V_*$ map of the data cube of NGC 3115. The black line represents the line of nodes of the map. Upper right: $\sigma_*$ map of the data cube of NGC 3115. Bottom: RGB composite image with the data cube of NGC 3115 (before starlight subtraction) collapsed along the spectral axis shown in green, the image of the red wing of the broad component of H$\alpha$ shown in red, and the $\sigma_*$ map shown in blue. In all maps, the kinematic center, which is coincident with the stellar bulge center, is shown with a black cross.\label{fig5}} 
\end{figure}

A careful analysis of the kinematic maps and of the RGB composite image in Figure~\ref{fig5} reveals that the area with the highest $\sigma_*$ values has an elliptic shape whose major axis is approximately perpendicular to the line connecting the $\sigma_*$ peak and the AGN we detected. We believe that this behavior, together with the fact that the $\sigma_*$ peak is not coincident with the stellar bulge center, may be the result of a perturbation in the stellar kinematics in the nuclear region of NGC 3115 caused by the black hole corresponding to the off-centered AGN. The value of $V_*$ in the spectrum of the off-centered AGN is $V_*(AGN) = 56.5 \pm 3.7$ km s$^{-1}$, while the value of the velocity corresponding to the peak of the narrow component of the H$\alpha$ emission line is $V_{H\alpha}(narrow) = 140.7 \pm 8.1$ km s$^{-1}$. This indicates that the stellar and the emission-line kinematics in the region of the AGN are not connected to each other.

After a galaxy merger, the supermassive black holes of the galaxies sink toward the center, via dynamical friction, form a binary and, then, when the emission of gravitational waves carries away the remaining angular momentum, the black holes coalesce (for more details, see Begelman et al. 1980; Merritt \& Milosavljevic 2005; Komossa 2006; Merritt 2006). The emission of gravitational waves is asymmetric and carries linear momentum. Then, to enforce global conservation of momentum, the system recoils, which results in a ``kick'' that displaces the merged remnant of the binary black hole coalescence (for more details, see Sundararajan et al. 2010, and references therein). Therefore, we propose that the off-centered black hole in NGC 3115 may be part of a binary system before the coalescence or it may also be the displaced merged black hole after the coalescence of the binary system. An analysis of the kinematics of the off-centered AGN around the stellar bulge center could help to decide between the two proposed scenarios. The redshift in the broad component of the H$\alpha$ emission line could, in principle, be used for such analysis. However, since this displaced broad component may be also the result of gravitational redshift, we believe that this analysis is not reliable and it will not be performed here.

The detection of an off-centered AGN in a data cube was originally done by \citet{med93} in an analysis of NGC 3227. In the case of NGC 3115, the detection of the off-centered AGN was only possible due to the methodologies we used during the data treatment and data analysis. This shows the importance of such procedures in the analysis of data cubes obtained with GMOS as well as with other instruments \citep{men14}.

\acknowledgments

This work is based on observations obtained at the Gemini Observatory. We thank FAPESP for support under grants 2012/02268-8 (RBM) and 2012/21350-7 (TVR) and also an anonymous referee for valuable comments about this Letter.

{\it Facilities:} \facility{Gemini:Gillett(GMOS)}.

\clearpage

\clearpage

\clearpage

\clearpage

\clearpage

\end{document}